\documentstyle[aps,prb,epsf]{revtex}
\begin{document}
\input psfig.sty
\draft

\twocolumn[\hsize\textwidth\columnwidth\hsize\csname @twocolumnfalse\endcsname

\title{First order valence transition in Ce and YbInCu$_4$ in the 
$(B,T)$ - plane}
\author{M. O. Dzero$^1$, L. P. Gor'kov$^{1,3}$, A. K. Zvezdin$^{1,2}$}
\address{$^1$ National High Magnetic Field Laboratory,
Florida State University, Tallahassee, FL, 32304, USA\\
$^2$ Institute of General Physics, Russian Academy of Sciences,
117942 Moscow, Russia\\
$^3$ L. D. Landau Institute for
Theoretical Physics, Russian Academy of Sciences, 117334 Moscow, Russia}
\date{\today}
\maketitle

\begin{abstract}
The puzzling properties of the first order phase transition in 
YbInCu$_4$ and its alloys in the wide range of magnetic fields and 
temperatures are perfectly described in terms of an entropy 
transition for free Yb ions. In particular, it turns out that the 
transition line in the $(B,T)$-plane is very close to the elliptic 
shape, as it has been observed experimentally. 
Similar calculations are done, and the experiments are proposed 
for the $(\gamma-\alpha)$ phase transition in Ce in 
Megagauss fields.
We speculate, that in case of YbInCu$_4$ the first order transition 
is a Mott transition between a higher temperature phase 
in which localized moments are stabilized by the entropy terms 
in the free energy, and a band-like non-magnetic ground state of 
the $f$-electrons. 
\end{abstract}
]
For years the isostructural $\gamma-\alpha$ transition in metallic 
Ce has been a classic example of the first order transition 
into a state with an intermediate
valence ($\gamma$ - Ce$^{3.06+}\rightarrow{~\alpha}$ - Ce$^{3.67+}$.
For the phase diagram of Ce, see \cite{Koskenmaki}. Change in the
valence state was judged by the change in the unit cell volume). 
Although a structural transition
in a crystalline matter possessing a critical point in 
the $(P,T)$-plane is of a great interest by itself, the 
discovery of a non-integer valence in
$\alpha$ - Ce has opened the field of the
so-called intermediate or mixed valence (MV) states in the rare earths and
actinides (both for elemental metals and intermetallic compounds).

The ($\gamma-\alpha$) transition in Ce takes place in the pressure range
$P{\sim}10{\div}20$kBar. Therefore, the isostructural transition 
in YbInCu$_4$ \cite{Felner} at $T_s{\sim}40$K and at ambient 
pressure \cite{Sarrao3}, looking akin to the Ce ``isomorphic'' 
transition, has attracted recently a lot of interest due to the 
possibility of study of that phenomena in much greater details. 
(The major experimental results are best 
summarized in \cite{Sarrao4,Cornelius,Sarrao5}).

In what follows we address the issue of the phase diagram
of Ce and YbInCu$_4$ in the $(B,T)$-plane, where $B$ is a magnetic field, 
$T$ is a temperature. Indeed, among many interesting results of
\cite{Cornelius,Sarrao5}, the most suprising one is the universality
of the first order transition line for YbInCu$_4$ and its alloys.
Namely, being expressed in the reduced variables $(B/B_{c0}, T/T_{v0})$
the transition line separating the high temperature phase (paramagnetic,
local moments) and the low temperature ``metallic'' phase  is a perfect
circle (where $T_{v0}$ is the structural transition temperature
in the absence of the magnetic field and $B_{c0}$ is the critical field
at T=0).We will show that these results are well
described in terms of an entropy first order phase transition between
the local $f$-moment phase and another phase probably of a
less ordinary nature. {\it An origin} of this phase however 
seems not to be important if this second phase is characterized 
by a larger energy scale. The same ideas applied to 
the $(\gamma-\alpha)$ transition in Ce predict similar 
behaviour in high magnetic fields with $B_{c0}{\sim}200$ Tesla. 
This is the field range achievable for  modern Megagauss magnetic 
field experiments \cite{Megagauss}.

The valence of Ce in the $\gamma$-phase is very close to the integer
$f$-occupancy (see, e.g. review \cite{Lawerence}), i.e.,
in the atomic configuration $(Xe+4f5d6s^2)$ \cite{Landau} all $d$-
and $s$-electrons of Ce go to the metallic bands. In accordance with
the Hund's rule the ionic ground state has the total angular
momentum $J=5/2$ which is split further in the cubic environment
into a ${\Gamma}_7$ doublet and a ${\Gamma}_8$ quartet
(in the Ce $\gamma$-phase ${\Gamma}_7$ lies below ${\Gamma}_8$).
Similarly, for Yb, the atomic configuration $(Xe+4f^{14}6s^2)$
results in the trivalent Yb$^{3+}$ ionic configuration for the
high temperature YbInCu$_4$-C15b phase, leading, to a localized 
$f$-hole. The $f^{13} (J=7/2)$ ground state is split by the crystal field
into a quartet (${\Gamma}_8$) and two doublets
(${\Gamma}_6$ and ${\Gamma}_7$). Inelastic neutron studies
\cite{Severing} at T$>{45}$K reveal the crystal field scheme
with ${\Gamma}_6$ and ${\Gamma}_7$ lying at 3.2 meV and 3.8meV
respectively above the ground state quartet ${\Gamma}_8$.

The first order transition line is determined by:
\begin{eqnarray}
F_U(B,T)&=&F_L(B,T).
\label{one}
\end{eqnarray}
In (\ref{one}) $F_L$ and $F_U$ stand for the free energies of the upper and
lower phases. The main assumption we use below is that the characteristic
energies governing the behaviour of the two phases differ significantly.
We denote these scales as $T_{K}^{U}$ and $T_{K}^{L}$, two effective
``Kondo temperatures'', in accordance with the existing tradition in
the experimental literature to plot data versus the isolated Kondo center
properties \cite{Sarrao3,Cornelius,Sarrao5} (for extensive discussion
of the theoretical results for the degenerate Anderson models and
the experimental results, see \cite{Schlottmann}).

For YbInCu$_4$ $T_{K}^{U}\simeq{25}K$ 
while $T_{K}^{L}\simeq{500K}$~\cite{Sarrao6,Immer}.
With $T_v$, the temperature of the ``valence transition'', for
Yb and its alloys lying in the range of $10-100K$ and 
$B_c\sim{50}$Tesla
\cite{Cornelius,Sarrao5}. The $F_L(B,T)$ in (\ref{one}) can
be taken as a constant, neglecting the magnetic susceptibility
term, while for the $F_U(B,T)$ with the 
trivalent Yb$^{3+}$ considered as a local free ion, one has:
\begin{eqnarray}
F_U(B,T)&=&E_0 - T{\cdot}S(B,T)
\label{two}
\end{eqnarray}
with the band energy $E_0$ being actually temperature independent below
$T_v$ (this assumption is discussed in more details below).
Correspondingly, the first order transition line in the $(B,T)$ plane
is given by the equation:
\begin{eqnarray}
T\cdot{S(B,T)}&=&const,
\label{three}
\end{eqnarray}
where the entropy is determined by the Yb$^{3+}$ multiplet structure only.

The magnetic susceptibility ${\chi}(T)$ of YbInCu$_4$ 
above $T_{v0}=42K$ follows
the Curie-Weiss law with an effective moment only negligible (by 5\%) smaller
than the whole $J=7/2$ ground state moment \cite{Sarrao5,Immer}. Thus, 
we first neglect the crystal splitting and write:
\begin{eqnarray}
T\cdot{S(B,T)}&=&-T\ln\left\{\sum\limits_{m=-J}^{J}
\exp\left(-\frac{g_J{\mu}_B{B}}{T}m\right)\right\},
\label{four}
\end{eqnarray}
where $g_J$ is a $g$-factor (for $J=7/2, g_J=8/7$). From (\ref{four})
the relation $a{\mu}_B{B_{c0}}=T_{v0}$ between the 
critical field $B_{c0}$ at $T=0$
and the value of the structural transition temperature $T_{v0}$ at zero
field, is of the form:
\begin{eqnarray}
g_J J\mu_B{B_{c0}}&=&T_{v0}\ln{(2J+1)},
\label{five}
\end{eqnarray}
which gives $a\approx{1.9}$ for $J=7/2$, 
a result which is remarkably close to the experimental 
value $a\simeq{1.8}$ \cite{Sarrao5}.
\begin{figure}[h]
\vspace{-0.5cm}
\centerline{\psfig{file=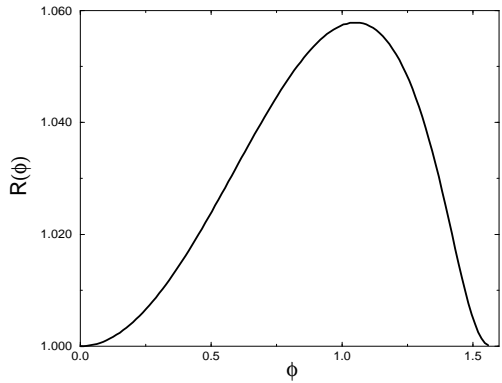,height=6cm,width=7.5cm,angle=-90}}
\caption{Function $R(\phi)$ (see text, Eqs.(9,10))
calculated for Yb$^{3+}$ (J=7/2). Deviations of $R(\phi)$ from
1 do not exceed 6\%.}
\end{figure}
We re-write the Eqs. (\ref{three},\ref{four})
for the phase transition line using the new reduced variables
$\beta=B/B_{c0}$ and $\tau=T/T_{v0}$ and with the help of 
(\ref{five}) we obtain:
\begin{eqnarray}
\tau&\ln&\left\{\sum\limits_{m=-J}^{J}
\exp\left[-m\left(\frac{\beta}{\tau{J}}
\ln{(2J+1)}\right)\right]\right\}=\nonumber\\
&=&\ln{(2J+1)}.
\label{six}
\end{eqnarray}
Using the parametric form $\beta/\tau{=}\tan{\phi}$ 
and the identity:
\begin{eqnarray}
\beta^2 + \tau^2&=&\tau^2\cos^{-2}\phi,
\label{eight}
\end{eqnarray}
one may re-write (\ref{six}) as:
\begin{eqnarray}
\beta^2 + \tau^2&=&R(\phi),
\label{nine}
\end{eqnarray}
where
\begin{eqnarray}
&&R(\phi)=\ln^2{(2J+1)}~\times\nonumber\\
&&\left\{\cos\phi\cdot\ln\left[\sum\limits_{m=-J}^{J}
\exp\left(\frac{-m\ln{(2J+1)}\tan\phi}{J}\right)\right]\right\}^{-2}.
\label{ten}
\end{eqnarray}
The plot of the function $R(\phi)$ is shown in Fig.1. 

Since the deviation of $R(\phi)-1$ from zero does not exceed $0.06$,
we arrive to the main result of \cite{Sarrao5,Immer}:
\begin{eqnarray}
\beta^2 + \tau^2 &{\simeq}& 1.
\label{eleven}
\end{eqnarray}
Postponing a detailed discussion for further publication, it is nevertheless
necessary to highlight some essential features of our picture. The first
interesting question is whether the account of crystal field split multiplets
would improve the overall agreement with experiments. Although we have
analyzed the relation:
\begin{eqnarray}
a\mu_{B}B_{c0}&=&T_{v0}
\label{twelve}
\end{eqnarray}
in terms of the crystal field Hamiltonian:
\begin{eqnarray}
\hat{H}&=&\hat{H}_{crystal} + g_J\mu_{B}\hat{\bf J}\cdot{\bf B}
\label{thirteen}
\end{eqnarray}
we do not stay on the results here, because (\ref{twelve}) must 
display some cubic anisotropy which was not experimentally 
studied yet (two components in
(\ref{thirteen}) do not commute with each other). We will limit 
ourselves with a comment that the energy levels' scheme for 
(\ref{thirteen}) follows straightforwardly from making use of 
the explicit wave functions
\cite{Lea} for the representations $\Gamma_6, \Gamma_7$ and $\Gamma_8$.
For the magnetic field, applied along the main cubic
axis, it turns out that the experimental value $a\simeq{1.8}$
\cite{Sarrao5,Immer} is again closely reproduced in such analysis.
We also would like to emphasize that the entropy:
\begin{eqnarray}
S(T_{v0}){\Rightarrow}\ln\left\{4 + 2\exp\left(-\frac{E_6}{T_{v0}}\right) +
2\exp\left(-\frac{E_7}{T_{v0}}\right)\right\},
\label{fourteen}
\end{eqnarray}
with $E_6, E_7$ taken from \cite{Severing}, is rather close to its
value $0.8\ln{8}$ as integrated through $T_{v0}$ (see 
\cite{Sarrao5,Sarrao6,Immer}) for YbInCu$_4$, which indirectly 
confirms the applicability of an isolated crystal field split hole 
state for Yb$^{3+}$ paramagnetic ion.

The Yb$^{3+}$ hole occupation in the high temperature state 
determined from Yb-L$_3$ X-ray absorbtion for the most compositions 
studied in \cite{Cornelius} turns out to be really close to 
the Yb$^{3+}$ trivalent state. This last fact,
however, does not preclude yet that the upper phase may have developed
pronounced Kondo effects with $T_K\simeq{25K}$, as e.g. is stated
in references \cite{Cornelius,Sarrao5,Immer}. On the other hand, it is not
clear whether the existing data show considerable deviations from the free
ion behavior for the upper phase. However, if it were so, (\ref{two})
would not have been correct at low temperatures. In such case, 
one may choose for $F_U(B,T)$ another expression, say, 
the exact solution for the exchange model or for the 
degenerate Anderson model. In the non-magnetic phase the scale,
$T_K\approx{500K}$ \cite{Immer} is rather large and one may neglect the
temperature dependence in $F_L(B,T)$ at temperatures below $50-100K$.
As for the conduction band electrons, typical energy scale for Ce would be
of order of $1$eV. Such a scale for YbInCu$_4$ and its alloys comprises
probably only $\sim{0.1}$eV, as discussed below.

It would be interesting to check, of course, whether the circular shape
(\ref{eleven}) of the transition line in the $(B,T)$-plane is indeed due to
the entropy transition in the {\it free ion scheme} 
of Eqs.(\ref{two},\ref{three}) or the result could be merely 
robust numerically.
Unfortunately, the Anderson model thermodynamics in high magnetic 
fields has been studied in the Coqblin-Schriffer model limit 
(the charge is fixed) for Ce $(J=5/2)$ but not for Yb $(J=7/2)$. 
Even for Ce, there are published 
results only for magnetization and specific heat 
(see in \cite{Schlottmann}). To obtain the free energy expressions, 
one would need to integrate these data
back, or solve the Bethe Anzats equations again.
\begin{figure}[h]
\vspace{-0.5cm}
\centerline{\psfig{file=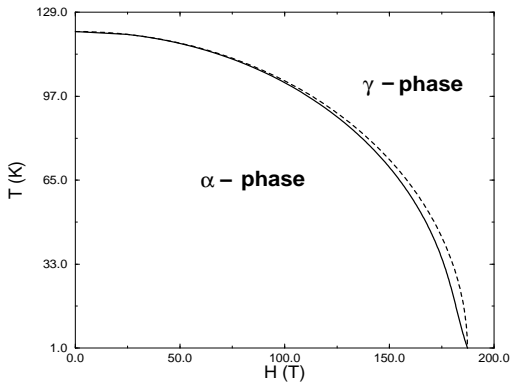,height=6cm,width=7.5cm,angle=-90}}
\caption{The line (solid) of the first order phase transition in the
$(B,T)$-plane for Ce $\gamma-\alpha$ transition. In the reduced
variables, deviations from the perfect circle (dashed line)
would not exceed 6\%.}
\end{figure}
In Fig. 2 the phase diagram in the $(B,T)$-plane shows the first order
phase transition line (solid), calculated for Ce according to
(\ref{four}) $(J=5/2, g_J=6/7)$. Its shape is again close to
a cirlce in reduced variables (dashed line): the deviations from the circle
do not exceed $\approx{8\%}$. The metamagnetic $(\gamma-\alpha)$ transition
in Ce has not been measured yet, to the best of our knowledge. 
One sees that the low temperature values of magnetic fields 
are in the experimentally accessible Megagauss range
($B^{\gamma}\sim{185}$ Tesla). The experiment presents a considerable
interest and allows one to verify the free ion model we
are using.

Finally, let us discuss the physics, which may be responsible for the
transitions in Ce and YbInCu$_4$. A first attempt to describe the
Ce $(\gamma-\alpha)$ transition was the Falicov-Kimball-Ramirez (FKR) 
model \cite{Kimbal}. Although the FKR model is capable to 
reproduce the appearence of the
critical point on the ($\gamma-\alpha$) transition line in the $(P,T)$- 
plane, it does not reproduce such crucial 
feature of the $\alpha$- phase as its intermediate valence.

So far, the ($\gamma-\alpha$) transition in Ce is usually 
interpreted in terms of the Kondo Volume Collapse (KVC) 
model \cite{Allen,Liu}. In the KVC model Ce atoms at the 
transition are treated as Ce$^{3+}$-ions in the both
$\alpha$ and $\gamma$ phases (approximately one electron 
in the $f$-shell), although in the two different Kondo regimes. 
As it is known, the Anderson impurity model
reproduces the Kondo behaviour in the regime when charge fluctuations
are fully suppressed, and provides for the $T_K$ the expression:
\begin{eqnarray}
T_K\propto\exp\left\{-\frac{\mid{\varepsilon_f^{*}}\mid}{\Gamma}\right\},
\label{fourteen2}
\end{eqnarray}
where $\mid{\varepsilon_f^{*}}\mid$ is the effective position of the
localized level below the chemical potential and the level's width
$\Gamma\propto{V^2\nu{(\epsilon_F)}}$
depends on the hybridization
matrix element,$V$, and the density of states at the Fermi level,
$\nu{(\epsilon_F)}$.

The KVC model \cite{Allen} connects the first order transition with
strong non-linear dependence of the Kondo scale (\ref{fourteen2})
($\mid{\varepsilon_f^{*}}\mid\ll{\Gamma}$) on the volume through
the volume dependence of the hybridization matrix element
(change in the volume of a unit cell 
is large, $\delta{v}/v\sim{20\%}$!). 
Our arguments in the beginning of the paper,
regarding the two different Kondo scales as needed for the 
applicability of (\ref{three}), agree well with the 
values \cite{Liu}:
\begin{eqnarray}
T_K^{\alpha}{\simeq}{81.2 \pm 12.2}{\text meV};
~T_K^{\gamma}{\simeq}{8.2 \pm 1.5}{\text meV}\nonumber
\end{eqnarray}

Nevertheless, the KVC model seems not to be applicable in case 
YbInCu$_4$, where the volume
changes are extremely small \cite{Sarrao3,Cornelius,Sarrao5,Sarrao6}.
On that reason, the FKR model has recently been 
revisited in \cite{Zlatic}. It is interesting, that although
being somewhat sensitive to the choice of the model parameters,
the elliptic shape of (11) for the phase transition line 
in the $(B,T)$-plane is preserved in the calculations \cite{Zlatic}. 
This is probably due to the same mechanism 
as above , i.e. due to large differencies in the
energy scales for the two phases (it seems however that the
constant $a$ in (12) strongly depends on the parameters choice).
Nevertheless, the FKR model can hardly be applicable for the
YbInCu$_4$ compound. In addition to its well known drawbacks, such
as an absence of hybridization, large changes in the $n_f$ occupancy, 
it seems that the peculiarities of this compound may originate
in somewhat unusual features of its non-magnetic analog LuInCu$_4$.
This point has already been discussed in \cite{Figuaora}. 
In \cite{Figuaora} the authors suggested a mechanism based on the
band structure calculations \cite{Takegahara,Monachesi} for the
semimetallic state observed both 
in LuInCu$_4$ and YbInCu$_4$ (Yb$^{3+}$, the
high temperature phase!)\cite{Cornelius,Sarrao5}. 
In this state $\varepsilon_{f}$ level falls into a ``pseudogap'' 
(or a dip in the DOS) at the choice of the chemical potential 
corresponding to the Yb$^{3+}$ configuration. 
If the exponential form for $T_K$ in (\ref{fourteen2}) remains 
correct, the strong non-linearity
in (\ref{fourteen2}) comes from the rapid changes in 
the {\it values of the DOS at the Fermi level}\cite{Figuaora}.
Thus, the high temperature state (the one above $T_{v0}\simeq{42}K$) 
is stabilized by the entropy gain, while the  phase
with the higher DOS at the Fermi level is prefered at
lower temperatures. The new state would,
hence, correspond to a non-integer valence Yb$^{2.8+}$ as measured in
\cite{Cornelius} to be Yb$^{2.8+}$ (an interaction term needs to 
be added to stabilize the valence). 

The feature, which remains not well understood with the above
explanation is that the Hall coefficient sharply decreases
in the low temperature state, as if the number
of carriers is getting comparable with the stoichiometric
value for the divalent Yb-ion (no hole in the $f^{14}$-shell).
The increase is too large compared
with the valence change $\approx{0.2}$ \cite{Cornelius}.
The energy scales involved and change  
in the Hall coefficient are also not consistent with the results of
\cite{Takegahara,Monachesi}. 
Indeed, $\Delta{E}$, the energy change per Yb
using Eqs. (\ref{one},\ref{two}) is:
\begin{eqnarray}
\Delta{E}&=&T_{v0}\ln{(2J+1)}\sim{90K}
\label{seventeen}
\end{eqnarray}
i. e. is too small to account for the large variation in the
number of carriers. We propose another view on this problem,
namely a weak Mott transition. At $T>{T_{v0}}$ localized moments
Yb$^{3+}$ are stable due to the entropy gain, 
and exist as the localized holes \cite{Kimbal}.
(Yb$^{3+}$)CuIn$_4$ is a band-like semimetal with a small
carriers concentration. Below $T_{v0}$ the valence change is
small because, unlike in the FKR model, only interactions
with small electron-hole pockets are essential. 
We speculate that after the transition into the
low temperature phase even the $f$-electrons form a band state, so that
a small change in ``occupation'' numbers does not contradict to
an emergence of a large $f$-like Fermi surface. An indirect support to
these views may be found in the recent band structure calculations
\cite{Antonov}. From \cite{Antonov}, one may conclude that the change
of the Yb valence $\sim{0.2}$ at the 42K would just result
in a shift  of the chemical potential by $\sim{0.01}$ eV inside the 
strongly featured DOS with a broader width of the order of $0.1$eV
(see Fig. 8 in \cite{Antonov}). This shift provides a correct 
magnitude for the energy change as estimated by (\ref{seventeen}).
The band picture is also in a reasonable agreement
with rather high value for the Sommerfeld 
coefficient in the linear term for the electronic
specific heat, $\gamma\approx{55}mJ/mole\cdot{K^2}$.

To summarize, we have shown that 
the entropy transition between the free ion
paramagnetic state and the low temperature metallic state
perfectly explains not only the elliptic shape of the 
transition line in the $(B,T)$-plane but also provides
correct numerical results for its parameters. 
Basing on similar calculations, we have suggested an
experiment on the metamagnetic transition between the
($\gamma-\alpha$) phases in cerrium. 
As for the true nature of the
transition itself, we suggest that in YbInCu$_4$ it is a weak 
Mott transition between a $f$-band metal and the 
semimetallic phase 
with the localized Yb$^{3+}$-holes.  

One of us (LPG) is grateful to Z. Fisk and J. Sarrao
for numerous stimulating discussions and for updating him 
with the recent references. He also thanks P. Schlottmann 
for the information regarding the exact numerical solutions
for the asymmetric Anderson model. H. Schneider-Muntau
has kindly provided us with references \cite{Megagauss} 
on the experiments with Megagauss fields.
A. K. Zvezdin thanks the NHMFL for the kind hospitality. 
This work was supported
by the National High Magnetic Field Laboratory through the NSF
cooperative agreement No. DMR-9527035 and the State of Florida.

\end{document}